# Transport properties of mid-infrared colloidal quantum dot films

Emmanuel Lhuillier, Sean Keuleyan, and Philippe Guyot-Sionnest
*James Franck Institute, 929 E. 57th Street, The University of Chicago, Chicago, Illinois 60637, USA*

**Abstract:** The transport and thermal properties of HgTe colloidal quantum dot films with cut-off wavelengths in the mid-IR are investigated. The cut-off wavelength of this material can be tuned over the 3-5 µm range, which makes it a promising alternative to existing high cost detectors. Post deposition processes such as ligand exchange and atomic layer deposition are investigated as a way to increase the carrier mobility.

## I - INTRODUCTION

Since the mid 90's colloidal quantum dots (CQDs) have seen growing interest[1]. Initially based on cadmium binary semiconductors[2], nanocrystals can now be grown from a wide range of materials which ensure a large tunability of their band-edge energy from UV to infrared (IR)[3]. Colloidal materials were of interest for applications in optoelectronics early on, and they have been successfully used for LEDs[4] and photovoltaic devices[5,6]. In vivo biological imaging has pushed the interest of these materials toward longer wavelengths. CQDs of the lead chalcogenides have reached as far into the IR as 0.30 eV (4.1µm) with PbSe[7] at low temperature.. Such a value is very close to the bulk value (0.27eV) and thus PbSe is not a good candidate to go further in the mid-IR. Semimetals, such as HgTe could provide a band-edge tunable through the full infrared range by quantum confinement and are of great interest in this aim. Here, we demonstrate the first photoconductive response to wavelengths as long as 7µm in CQD based devices.

The 3-5 µm wavelength range is of particular importance for imaging since (i) it overlaps with the atmospheric transparency window. (ii) In this range of wavelength self-emission starts to prevail over reflected/scattered light. Achieving CQD based devices operating through this range of wavelength creates a new paradigm for low-cost mid-IR detection. In this paper we report optoelectronic properties of mid-IR devices based on HgTe CQDs. After a short introduction on the synthesis of the material, we present the obtained performances with a focus on transport properties. The last section of the paper reports our latest progress on the surface chemistry toward controlling the interdot matrix, typically the organic molecules used to passivate the particle surfaces.

## II - MATERIAL SYNTHESIS

The synthesis of HgTe CQDs has been previously reported by several groups, but most of the results are limited to wavelengths in the near-IR[8910]. Heiss and coworkers achieved particle sizes for a band edge near 3µm. Diameters above 4 nm required a heat treatment which significantly broadened the size distribution. [11]. Our initial synthesis was based on the reaction of mercury acetate in ethanol or butanol with elemental tellurium dissolved in n-trioctylphosphine (TOP)., see references 12 and 13. This gave particles with diameters larger than 10 nm with cut-off wavelengths in the mid-IR, up to 5µm at room temperature, see Figure 1 (b). The particles were found to be aggregated as shown by the transmission electronic microscopy (TEM) image in Figure 1 (a). Such aggregation is confirmed to be present in solution as well by dynamic light scattering (DLS), with aggregate size in the several hundred nanometer range.

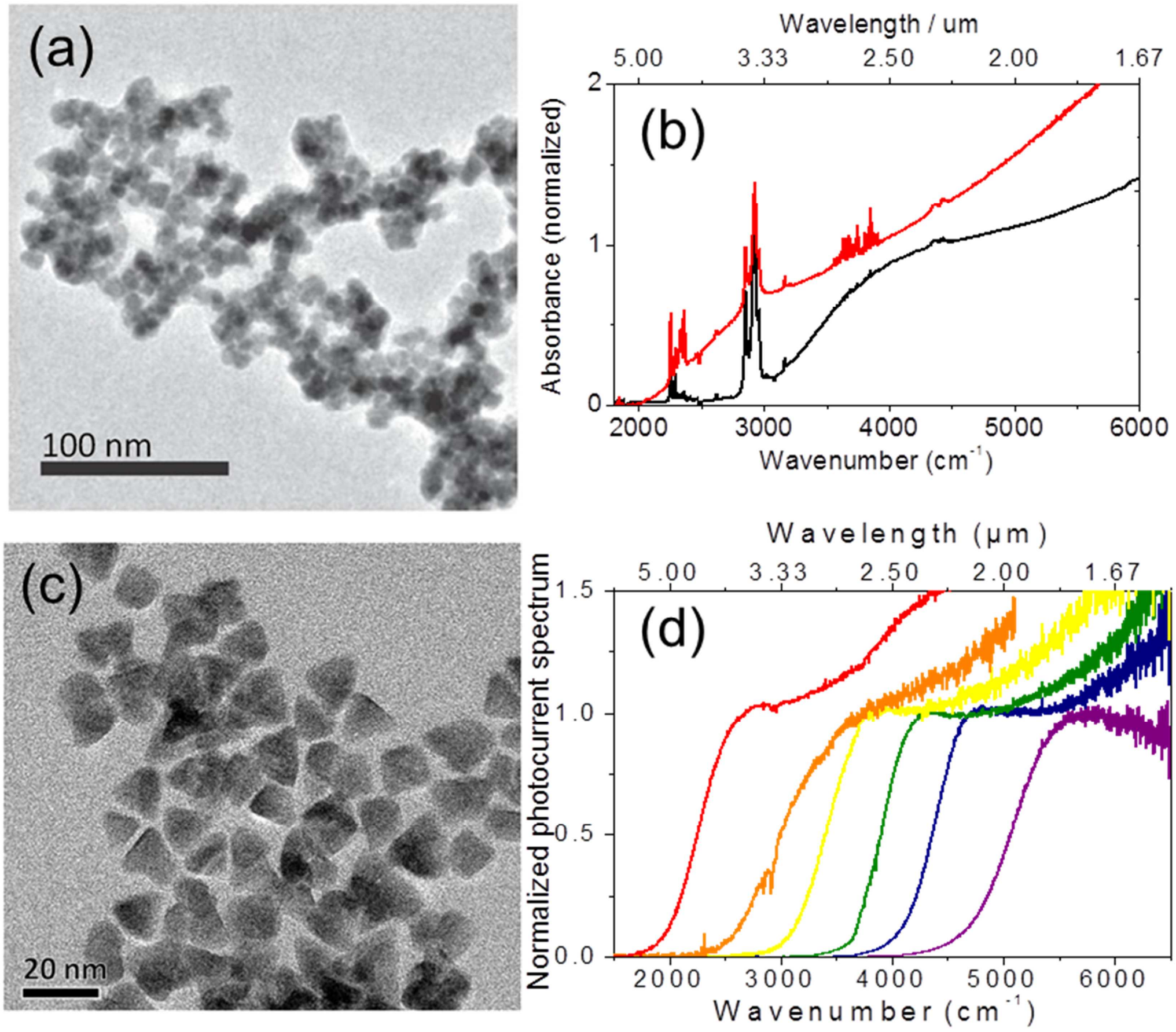

*Figure 1 (a) Transmission electronic microscopy image of the first generation of CQD. (b) Absorption spectra of the CQD of the first generation. (c) Transmission electronic microscopy image of the second generation of CQD. (b) Photocurrent spectra of the CQD of the second generation*

To achieve better size distributions for sharper optical responses, an improved synthesis was developed.[14] In this method, mercury(II) chloride is first dissolved in oleylamine, and tellurium in TOP is then quickly injected. This method leads to particles with tetrahedron like shapes, see Figure 1 (c). Improved size distributions give sharper ensemble spectra (see the photocurrent spectra in Figure 1 (d)), revealing excitonic peaks corresponding to the lowest energy transition, as well as a second feature at higher energy. In contrast to the previous method, the particles are not aggregated, which allows for better access of the particle surfaces for further for processing.The particle size is conveniently controlled by the reaction temperature and duration.

In all cases, after the synthesis, CQD growth is quenched using thiols and redispersion in tetrachloroethylene (TCE). To measure the photoconductive response, we prepare thin films of CQDs. The particles, initially in TCE, are precipitated by adding alcohol and centrifuging to remove excess reagents and ligands. The obtained solid is then dried under $N_2$ flow and redispersed in a 9:1 (v/v) hexane-octane mixture. This solution is then dropcast onto interdigitated electrodes (50 pairs of 5mm long Pt electrodes with a 10µm spacing). The sample is then mounted on the cold finger of a closed cycle helium cryostat. Current measurements are performed using a picoammeter Keithley 6487. FTIR measurements are made using a NICOLET magna IR 550 FTIR, except for Figure 1 (a) where a home-made FTIR was used. Noise measurements are made using a femto DLCPA 200 amplifier (also use as bias source) and a Stanford instrument SR 760 spectrum analyzer for the data acquisition.

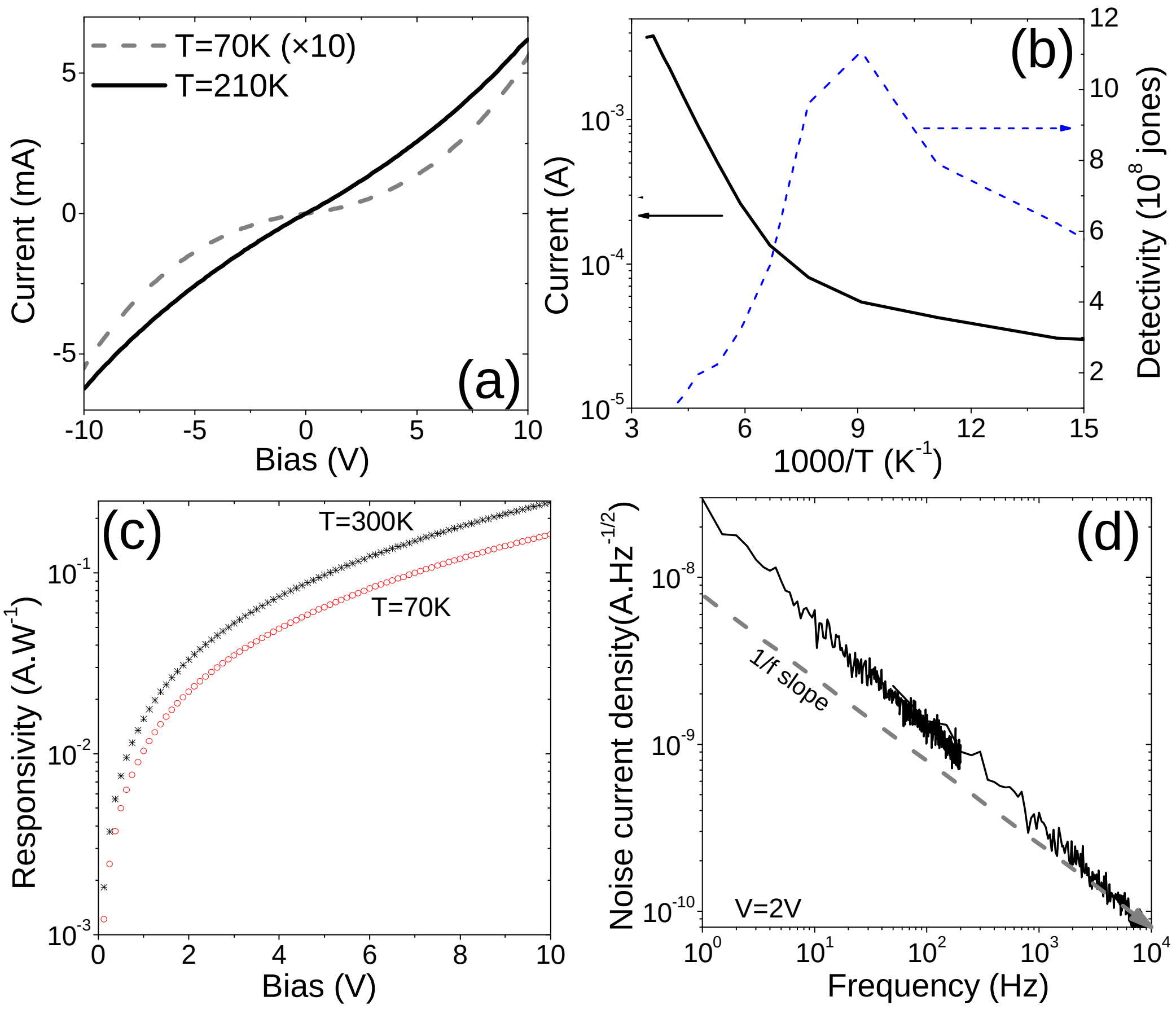

*Figure 2(a) I-V curve of a 3µm cut off wavelength sample at two different temperatures. (b) Dark current and detectivity as a function of the inverse of the temperature, under 2V (5V for the detectivity curve). (c) Responsivity as a function of the applied bias for two operating temperatures. (d) Noise current density as a function of the signal frequency under a 2 V bias at room temperature.*

## III – PERFORMANCE OF EARLY MATERIAL

### *A – TRANSPORT PROPERTIES*

I-V curves measurements between 3 and 300 K exhibit ohmic behavior at high temperature, deviating from linearity as the temperature is decreased. The thermal dependence of the material exhibits an Arrhenius behavior in the 150-300 K range, with an activation energy close to half of the optical band-edge energy. This shows that the transport is limited by thermal generation of carriers as for intrinsic semiconductors. At lower temperature, the activation energy is considerably reduced which indicates the presence of an impurity level close to either the valence or conduction band.

At room temperature, from the dark conductivity and a Boltzmann population, the carrier density is roughly $10^{-3}$carriers per dot, assuming a density of states equal to two. For a film a few hundred nanometers thick, we estimate the mobility to be in the $10^{-2}$-$10^{-1}$ $cm^2V^{-1}s^{-1}$ range. These values are large for unprocessed CQDs and this is certainly linked to the aggregation of the particles seen in the first generation material. Indeed the second generation leads to lower values of the mobility, typically a decade lower.

*B – OPTICAL PROPERTIES*

The band edge energy of the particle is not only dependent on the particle size (roughly as $R^{-\Delta}$, with $\Delta$ between one and two), but also on the temperature. HgTe presents a positive value for $\frac{dE_G}{dT}$, which leads to an increase in the cut-off wavelength of the sample as the temperature is decreased. Thus a sample with a cut-off wavelength at 5µm at room temperature has a 7µm band edge at 80K, see Figure 3.

Measured values for $\frac{dE_G}{dT}$ are equal to $300\mu eV.K^{-1} \pm 100\mu eV.K^{-1}$ which is lower than the value for the bulk ($530\ \mu eVK^{-1}$, see reference 15).

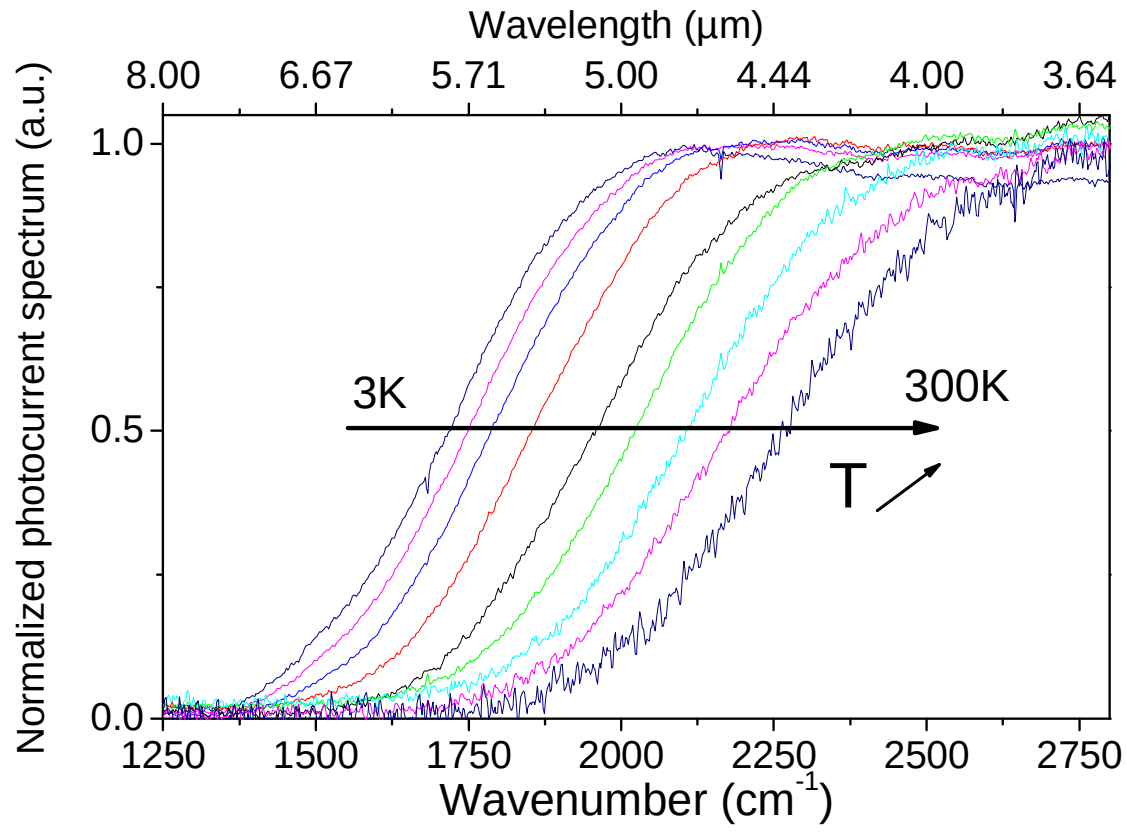

*Figure 3 Normalized photocurrent spectra at different temperatures from 3K to 300K, under 10V.*

For Stransky-Krastanov (epitaxially) grown quantum dots (QDs) used as infrared photodetectors[16] it is usually observed that the low density of QDs, typically a few $10^{11}cm^{-2}$, is a significant hurdle to achieving large quantum efficiency. In CQD, due to the higher packing density, we can achieve a larger absorption with thinner films. We have deposited films of CQDs on silicon wafers and measured their thickness by ellipsometry and atomic force microscopy. Then their absorption coefficients are determined, using FTIR, see Figure 4 (a). The obtained absorption depths are shorter than 1µm, which is similar to the values reported for bulk HgCdTe[17,18,19] diodes in the 3-5µm range. Moreover the comparison of the experimental data with the simulation, performed using a simple two band k·p code, leads to fair agreement, see on Figure 4 (b).

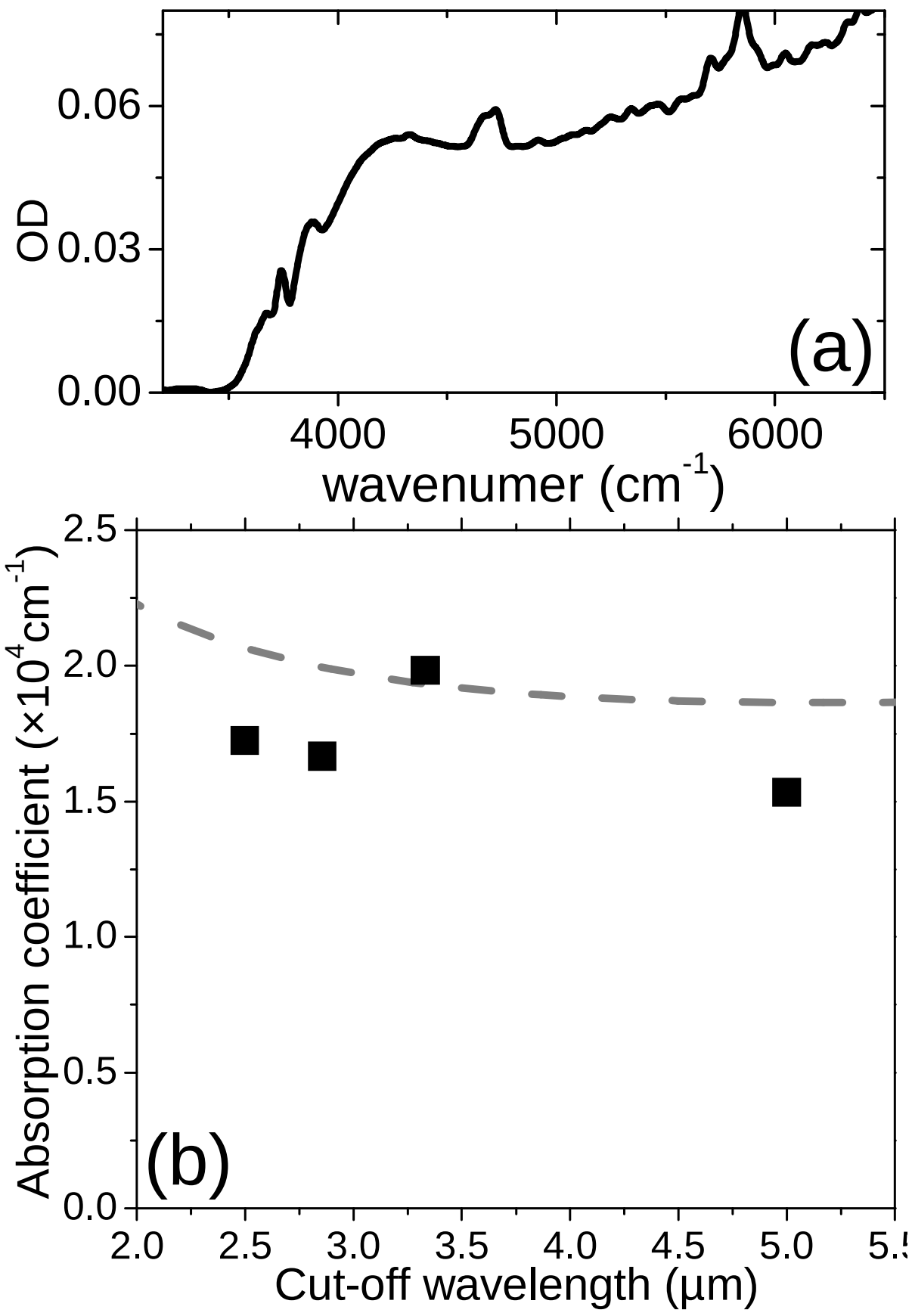

*Figure 4 (a) Absorption spectrum of a thin film (thickness≈60nm). Absorption coefficient as a function of the cut-off wavelength for several films. The line is obtained through a k·p simulation for comparison.*

*C – DETECTOR PERFORMANCES*

The values of responsivity achieved with this material were as high as $0.25A.W^{-1}$ at room temperature with little processing. For comparison the responsivity as a function of the cut-off or peak wavelength value are reported on Figure 5. External quantum efficiency are close to 10% at room temperature. The best commercial devices achieve almost 100% yield but require low operating temperatures to achieve such performances. There are a number of avenues available toward improving the responsivities of the HgTe CQD based devices and these are discussed in the next section.

The main drawback of colloidal materials is the large 1/f noise[13] which is observed over the quite broad range of temperatures 100K-300K. This currently limits the detectivity of these devices to the range of $10^8$-$10^9$ Jones.

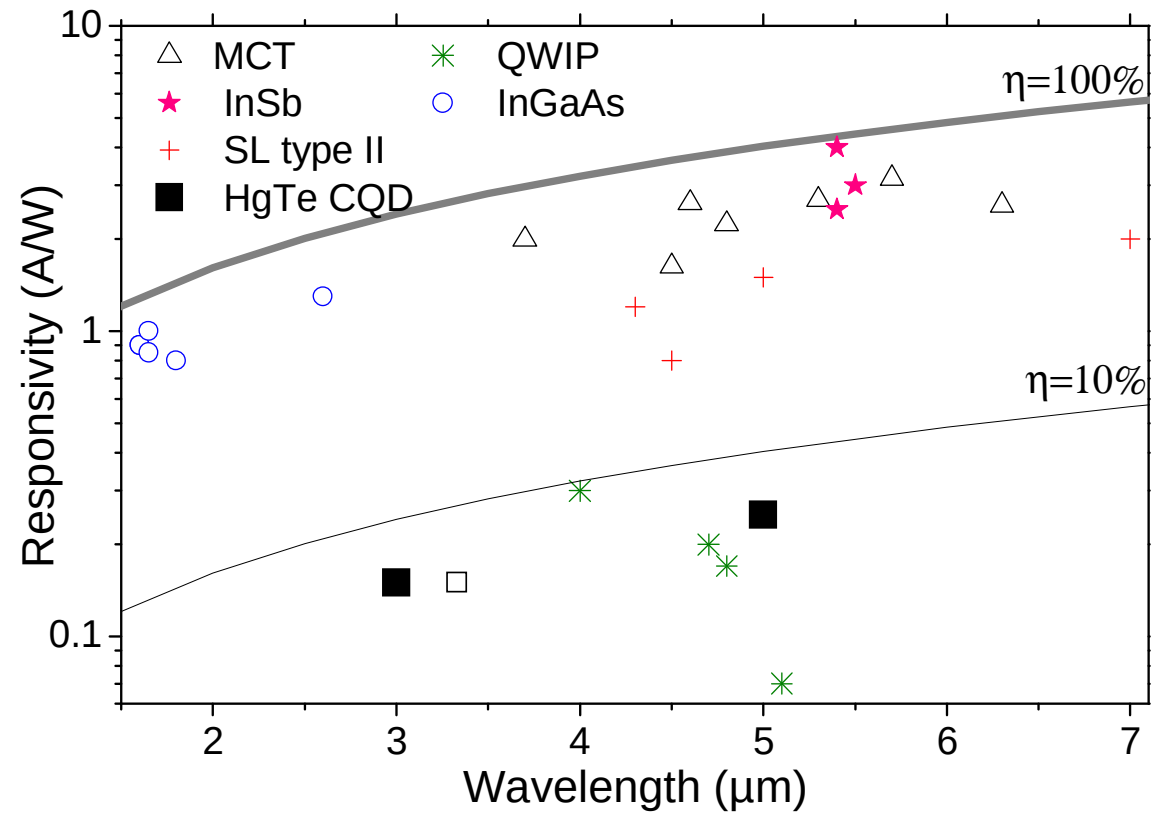

*Figure 5 Responsivity as a function of the cut-off (or peak) wavelength value for different technology over the near and mid IR.*

## IV – OPTIMIZING TRANSPORT PROPERTIES

### *A – COLLOIDAL MATERIALS IN INFRARED*

For colloidal materials, the overall conductivity remains low and this weakness can be attributed to the presence of organic ligands in the film. These ligands are required for the controlled growth and colloidal stability of particles in solution, as well as passivating the particle surface. Once processed into a film, the organic ligands act as an insulating barrier to charge transport, leading to the low carrier mobilities. In addition, organic molecules have vibrational transitions with energies similar to the bandgap of the CQDs, enabling energy transfer, reducing the exciton lifetime.[20].

### *B – ATOMIC LAYER DEPOSTION PROCESSING*

A possible way to process CQD films is the deposition of an inorganic matrix using atomic layer deposition (ALD). ALD of ZnO on films of CdS CQDs has been reported.[21] This led to a large increase in carrier mobility and a faster time response. Moreover, embedding CQDs in a matrix provides better surface passivation[21] [22] and may improve the absorption coefficient. The introduction of an inorganic matrix increases the dielectric constant of the surrounding reducing dielectric screening of the electromagnetic field, through the screening factor[23] $\left(\frac{3\varepsilon_{medium}}{2\varepsilon_{medium}+\varepsilon_{dot}}\right)^2$.

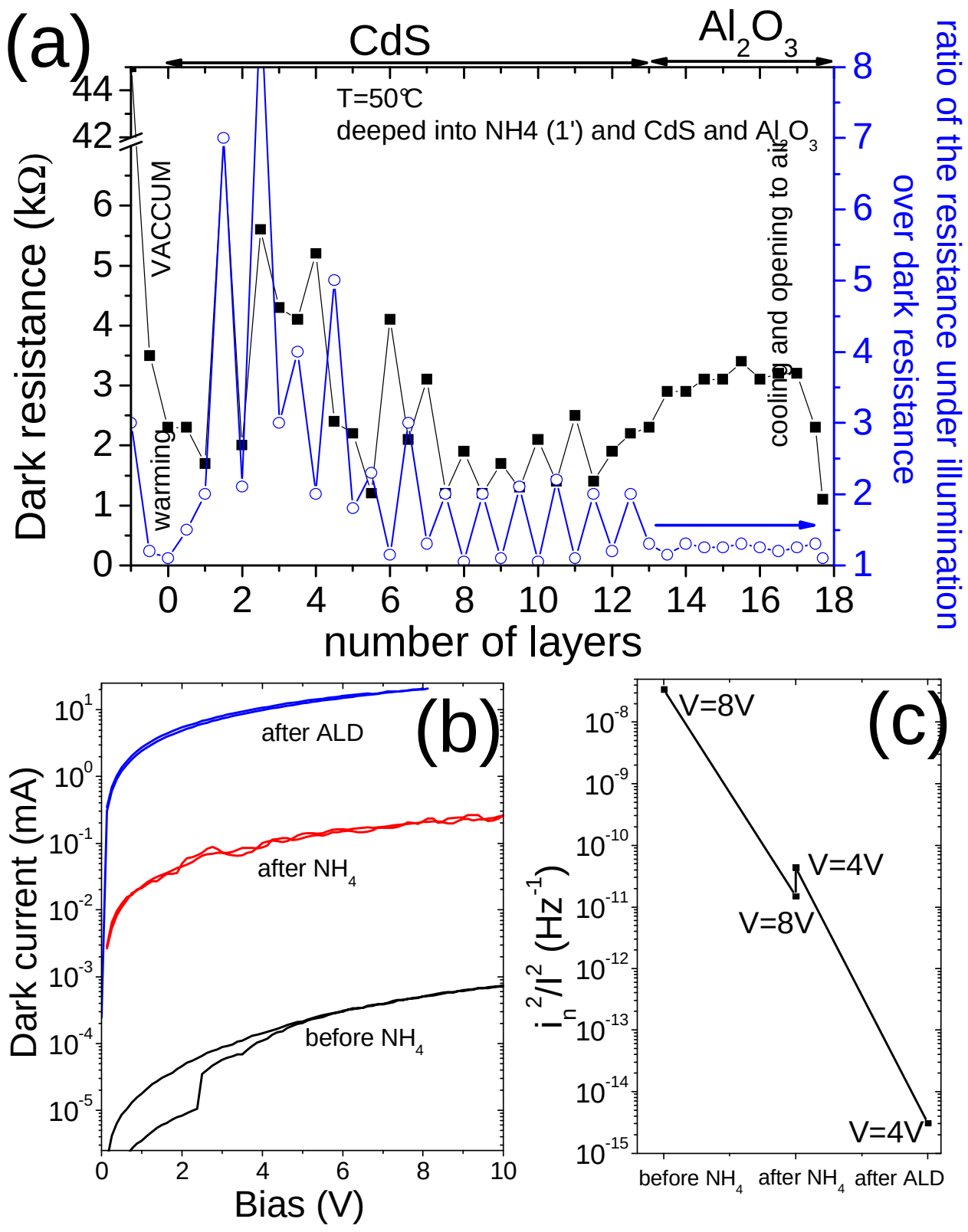

*Figure 6 (a) Evolution of the dark resistance and ratio of the resistance under illumination over the dark resistance during the atomic layer deposition process. (b) I-V curves for the same sample, before*

*and after the $NH_4OH$ cleaning, and after the ALD process. (c) Evolution of the ratio of the square of the noise over the square of the current after $NH_4$ treatment and the ALD process.*

The films are processed as follow: they are first dipped in warm $NH_4$ solution (50°C) in order to remove the excess of organic (see next section IV-C for the effect of this step). Then the film is introduced in the vacuum chamber of the ALD setup. The chamber is pumped up to a one torr pressure and the sample is heated to 50°C. Then we proceed to deposit CdS (using dimethyl cadmium and $H_2S$) or/and $Al_2O_3$ (with trimethyl aluminum and water). The change of conductivity during the process can be followed on Figure 6 (a). We observe that the conductivity rises with the removal of the organic ligand but also after the ALD process, see Figure 6 (b). This rise of the conductivity also comes with a better noise over current ratio (square of the ratio of the noise current density over the dark current), as shown on Figure 6 (c). Nevertheless in spite of the improved conductivity, the photo response is reduced and the sample is blind after the ALD process. Thus, up to now, ALD appears as an interesting process to tune the dark conductivity but decrease the photoconductivity properties. Similar result has been obtained by depositing only CdS or by depositing CdS and thiol ligand.

*C – LIGAND EXCHANGE PROCESSING*

As an alternative to ALD we investigate the ligand exchange process to tune the conduction properties of the CQD film. Ligand exchange is a suitable method to replace the initial long alkanes with shorter ones. Molecules like ethanedithiol or -diamine reduce the interparticle distance, increasing carrier mobilities. This process comes with a reduction in film volume[24] and the formation of cracks[25], see the inset of Figure 7(a).

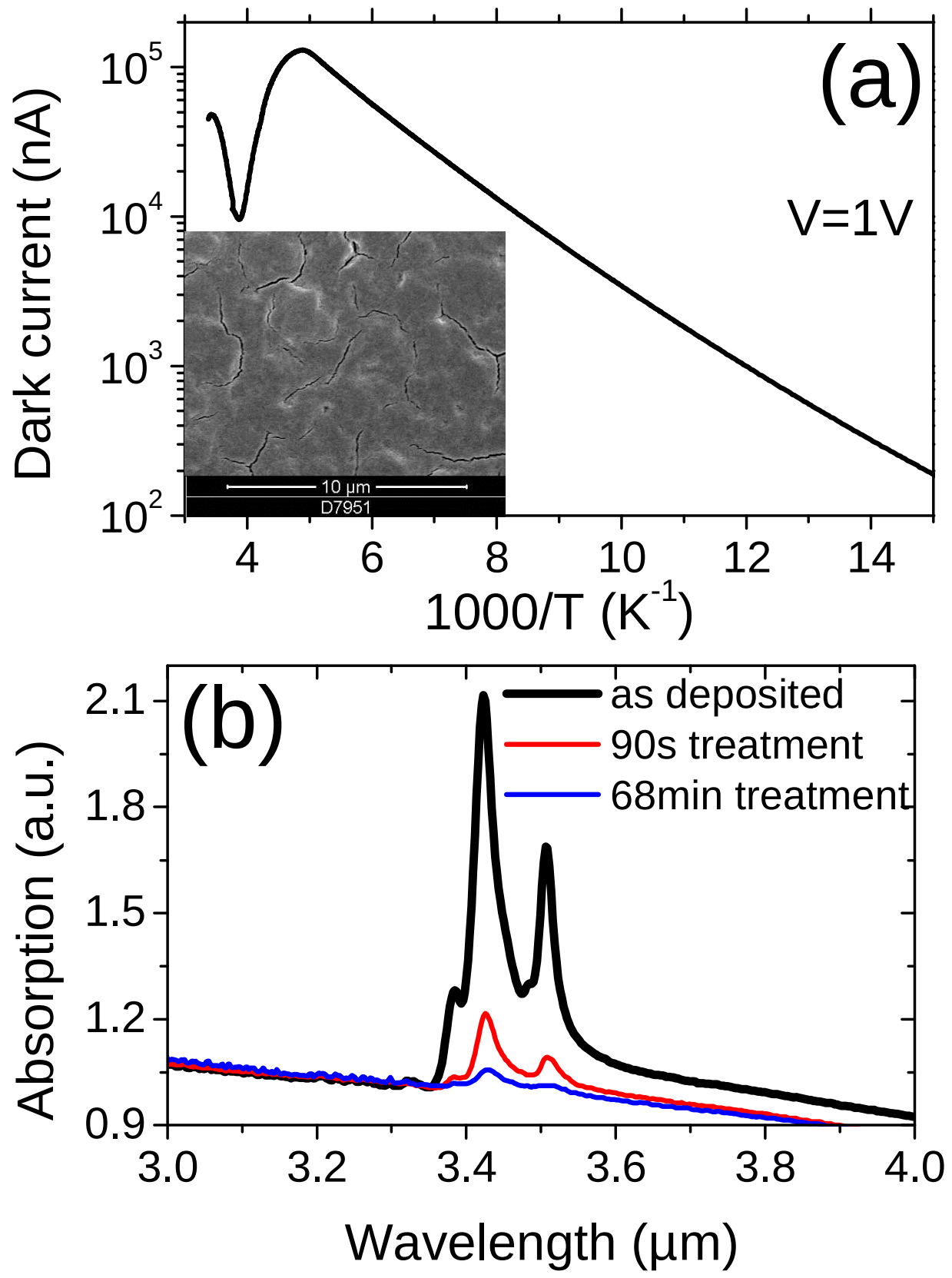

*Figure 7 (a) Dark current as a function of the temperature for a sample processed with ethanedithiol with a 3µm cut-off wavelength, under 1V bias. The inset is a scanning electronic microscopy picture of a film after treatment with ethanedithiol, showing the presence of cracks (b) Change in the magnitude of the CH absorption for different times of dipping the film in $NH_4OH$ in ethanol solution. The baseline level of each graph is taken equal to one .*

In narrow band gap materials, like HgTe, the introduction of charges even at the surface of the particle can easily shift the absolute energy of the Fermi level. Thus we observe that contrary to the raw material which presents an Arrhenius behavior, The *I-T* curve of films processed with ligand exchange is no longer monotonic, see Figure 7 (a). Non-monotonic behavior of the temperature dependence of the conductance has already been reported in bulk HgTe and attributed to acceptor states in the conduction band. [26] These acceptor states are localized just above the bottom of the conduction band (between 1 and 10meV above). In a disordered system like HgTe CQD we believe that the polydisperisty in size will hide this low energy resonance.

Removal of the excess ligand can also be achieved by dipping the film in a solution of ammonium hydroxide in ethanol, see Figure 7 (b). The ammonium hydroxide solution decreases the magnitude of the C-H peak, without affecting the excitonic feature as long as the solution is not too concentrated.

## V – CONCLUSION

To conclude we present HgTe CQDs with a band edge tunable to 7µm. This material, once processed into films presents photoconductive properties with 10% yield at room temperature. *1/f* noise is also observed for the considered range of temperature. We investigate two methods toward improving the conductivity. ALD leads to a huge tuning of the film conductivity and can reduce the relative importance of the noise. Ligand exchange is efficient to increase the inter-particle coupling and thus the carrier mobility. Further investigations are still required.

## VI - REFERENCES